\documentclass[12pt]{elsarticle}
\usepackage{mathtools}
\usepackage{graphicx}
\usepackage{float}
\usepackage{wrapfig}
\usepackage{rotating}
\usepackage[normalem]{ulem}
\usepackage{amsthm}
\usepackage{amsmath}
\usepackage{textcomp}
\usepackage{marvosym}
\usepackage{wasysym}
\usepackage{amssymb}
\usepackage[colorlinks,linkcolor=gray]{hyperref}
\usepackage{color}
\usepackage{enumitem}
\usepackage{tikz}
\usetikzlibrary{positioning,backgrounds,calc,shadings}
\usetikzlibrary{shapes,shapes.arrows,shapes.symbols,shadows}
\usetikzlibrary{arrows, decorations.markings,intersections}
\usetikzlibrary{patterns,arrows,decorations.pathreplacing}
\usepackage{pgfplots}
\usepackage{balance}
\usepackage{lipsum}
\usepackage[compact]{}
\usepackage{dblfloatfix}
\usepackage{float}
\usepackage{url}
\usepackage{listings}
\usepackage[justification=centering]{caption}
\usepackage{multirow}
\usepackage{adjustbox}
\usepackage{algorithm}
\usepackage{algorithmic}

\newtheorem*{th:rank}{Theorem \ref{th:rank}}
\newtheorem*{th:ind}{Theorem \ref{th:independence}}
\newtheorem*{th:low}{Theorem \ref{th:lowerbounds}}
\newtheorem{theorem}{Theorem}% [section]  

\newtheorem{lemma}{Lemma}

\theoremstyle{definition}
\newtheorem{definition}{Definition}% [section] 

\begin{document}
\title{\textbf{Quantum Algorithms for Identifying Hidden Strings with Applications to Matroid Problems}}
\author{Xiaowei Huang, Shihao Zhang and Lvzhou Li\footnote{L. Li is the corresponding author (lilvzh@mail.sysu.edu.cn)}}
\address{Institute of Quantum Computing and Computer Theory,  School of Computer Science  and Engineering, Sun Yat-sen University, Guangzhou 510006, China}
%\author{Lzayor}
%\ead{huangxw55@mail2.sysu.edu.cn}
\date{\today}

\begin{abstract}
In this paper, we explore quantum speedups for the problem, inspired by  matroid theory, of identifying  
a pair of  $n$-bit binary strings that are promised to have the same number of 1s and differ  in exactly two bits, by using  the max
inner product oracle and the sub-set oracle.     More specifically, given two string $s, s'\in\{0, 1\}^n$ satisfying the above constraints, for any $x\in\{0, 1\}^n$ the max
inner product oracle $O_{max}(x)$ returns the max value between $s\cdot x$ and $s'\cdot x$, and the sub-set oracle  $O_{sub}(x)$ indicates whether the index set of the 1s in $x$ is a subset of that in  $s$ or $s'$. We present a quantum algorithm consuming $O(1)$ queries to the max
inner product oracle for identifying the pair $\{s, s'\}$, and prove that any  classical algorithm requires $\Omega(n/\log_{2}n)$ queries. Also, we present a quantum 
algorithm consuming $\frac{n}{2}+O(\sqrt{n})$ queries to the subset oracle,  and prove that any  classical algorithm requires  at least  $n+\Omega(1)$ queries. 
Therefore, quantum speedups are revealed  in the two  oracle models. 
Furthermore, the above results are applied  to the problem in matroid theory of finding  all the bases of a 2-bases matroid, where a matroid
is called $k$-bases  if it has $k$ bases.

\end{abstract}

\maketitle

\section{Introduction}
%\subsection{Background}
The study of the string problem is important in theoretical computer science,
and has a wide range of applications in many fields such as
bioinformatics, text processing, artificial intelligence, and data mining,  etc.
Since the 1960s, the string problem has received much attention and  been
extensively studied in classical computing with a series of efficient 
algorithms  proposed,  including string  matching\cite{DBLP:journals/siamcomp/KnuthMP77,DBLP:journals/jal/Colussi94}, string 
finding\cite{DBLP:journals/cacm/BoyerM77,DBLP:journals/ipl/Moller-NielsenS84}, string
learning\cite{DBLP:conf/focs/MargaritisS95,DBLP:journals/jcb/SkienaS95} and so on. Compared to conventional classical computation, a new computing paradigm called  \textit{quantum  computation} was proposed in  the 1980s   \cite{manin1980computable,IJOTP/Feynman82} and has developed rapidly 
\cite{RSPA/Deutsch85,RSPA/Deutsch92,
DBLP:conf/focs/Simon94,DBLP:journals/siamcomp/BernsteinV97,
DBLP:conf/focs/Shor94,DBLP:conf/stoc/Grover96,Zhang2022}.
Finding more problems that admit quantum advantage has become 
one of the focus issues in the field of quantum computing, naturally including  
the string problem. In fact, 
the study of quantum algorithms for string problems started very early. As a notable example, 
the Bernstein-Vazirani algorithm proposed in  1997 \cite{DBLP:journals/siamcomp/BernsteinV97} is
one of the most outstanding works, which can learn 
an $n$-bit hidden string using only one query to the inner product oracle, while the best deterministic classical algorithm needs $n$ queries to solve this problem. To date, a series of contributions have been made to solving string problems in the quantum setting  \cite{DBLP:journals/jda/HariharanV03,
DBLP:journals/algorithmica/Montanaro17,DBLP:conf/innovations/GallS22,
DBLP:conf/soda/AkmalJ22,DBLP:conf/swat/CleveIGNTTY12}.

String learning is one of the well-known research directions in string problems, which has
important applications in bioinformatics \cite{2013DNA}, data mining \cite{2012mining} and network security  \cite{2022AES}.
In general, string learning is such a problem that the goal is to
reconstruct a secret string hidden by an oracle through querying the oracle as few as possible. Considerable  effort has been devoted to  string learning in both classical and quantum computing over the past decades, involving different oracles and 
application scenarios.
% classical 
Skiena and Sundaram \cite{DBLP:journals/jcb/SkienaS95} showed that
$(\alpha-1)n+\Theta(\alpha\sqrt{n})$ queries to the sub-string oracle are sufficient to 
reconstruct an unknown string, with $\alpha$ being the alphabet size and $n$ being the length of 
the string, and applied their algorithm to reconstruct DNA sequences. 
Du and Hwang \cite{ding-zhu/hwang:2000} considered the subset \textbf{OR}
oracle in combinatorial group testing.
%quantum
Cleve \textit{et al.} 
\cite{DBLP:conf/swat/CleveIGNTTY12} gave a quantum algorithm using $\frac{3}{4}n+o(n)$
queries to the sub-string oracle to reconstruct an unknown string of size $n$. Dam 
\cite{DBLP:conf/focs/Dam98} gave a quantum algorithm framework to identify a secret
string of size $n$ using $\frac{n}{2}+O(\sqrt{n})$ queries to the binary oracle. 
Also, quantum algorithms associated with  the balance oracle and the subset \textbf{OR}
oracle were  proposed for the  quantum counterfeit coin problem
\cite{DBLP:journals/tcs/IwamaNRT12} and  combinatorial group testing 
\cite{DBLP:journals/qic/AmbainisM14}, respectively.
 Recently, Xu \textit{et al.}  \cite{xu2022quantum} gave a quantum algorithm using
$\lfloor n/2\rfloor$ queries to the longest common prefix oracle to learn a secret 
string of size of $n$, which has a double speedup over classical counterparts.
Another interesting work is to play Mastermind on quantum computers. 
Li \textit{et al.} \cite{li2022winning} studied quantum strategies
for playing Mastermind  with $n$ positions and $k$ colors using the black(-white)-peg oracle. Note these string learning problems are all about learning a \textit{single}  hidden string, and a natural 
generalization is to think about learning  \textit{multiple} hidden  strings for  revealing possible quantum  advantages.

In this paper, we investigate  quantum  algorithms to  learn a pair of binary strings which are almost  the same, that is, they have the same number of 1s and differ only in two bits. More generally, these two  constraints are ubiquitous in  a mathematic concept called  \textit{matroid}~\cite{JHUP/whitney35}. For example, any matroid with
more than 1 bases  has such a pair of bases that satisfy the above
condition. Therefore, an effective quantum algorithm  which learns a pair of binary strings that
are almost the same implies that it can find the bases of 
a matroid with two bases. In the following, we first introduce some related background knowledge for convenience, and then introduce our main contributions of this paper.

\subsection{Background Knowledge}

\noindent\textbf{Notations}. $[n]$ denotes the set $\{1,2,\cdots,n\}$. 
For $j\in[n],x\in\{0,1\}^n$, $x_j$ denotes the $j$-th bit of $x$. 
$|x|=\sum_{l=1}^{n}x_l$ is  the number of $1$s in $x$, i.e., the Hamming weight of $x$. 
 $e_j$ represents a string such that its $j$-th bit is $1$ and the other bits are $0$.  The symbols $\wedge$, $\vee$ and  $\oplus$  represent the logical operation \textbf{AND}, \textbf{OR}, and \textbf{XOR} (modulo-2 addition), and would act as corresponding bitwise operations when  applied to two bit strings.  The inner product  of  $x, y\in\{0,1\}^n$ is $x\cdot y=\sum_{j=1}^{n}x_{j}y_{j}$.

\begin{definition}[Max Inner Product Oracle]\label{def:rank-oracle}
     The max inner product oracle associated with a string subset  $S\subseteq\{0,1\}^n$ is a function 
    $ O^S_{max}:\{0,1\}^n\rightarrow\mathbb{Z}$ defined by 
    \begin{equation}
    \label{O_max_S}    O^S_{max}(x)=\max\{x\cdot s:s\in S\},
    \end{equation}
    for any $x\in\{0,1\}^n$. 
\end{definition}

\begin{definition}[Sub-set Oracle]\label{def:sub-string-oracle}
   A sub-set oracle associated with a string subset $S\subseteq\{0,1\}^n$  is a function 
    $ O^S_{sub}:\{0,1\}^n\rightarrow\{0,1\}$ defined by 
\begin{equation}
\label{O_sub_S} 
     O^S_{sub}(x)=\left\{
      \begin{array}{ll}
        1, \;\;\text{if}\; \exists\;s\in S\;\text{s.t.}\; \text{Idx}(x)\subseteq\;\text{Idx}(s),\\
        0, \;\;\text{otherwise}.
      \end{array}
      \right.
\end{equation}
for any $x\in\{0,1\}^n$,  where the function  $\text{Idx}(z)=\{j\in[n]:z_j=1\}$ represents the index  set of the 1s in any string $z\in\{0,1\}^n$. Then  $O^S_{sub}(x)$ indicates whether the index set determined by $x$ is a subset of any index set determined by $s\in S$.
\end{definition}

%---------------------------------------------------------
Now   a natural problem is: Given the oracle $O^S_{max}$ or $O^S_{sub}$, how can we identify
$S$ by using as few queries to the oracle as possible? In this paper, we will consider
a particular case of this  problem with a promise  inspired by matroid theory as 
follows.\\

\noindent\textbf{Hidden String Problem(HSP):} 
Given the oracle $O^S_{max}$ in Eq.~ \eqref{O_max_S} or $O^S_{sub}$ in Eq.~\eqref{O_sub_S}  with the promise that $S$ consists of two $n$-bit strings $s,s'\in\{0,1\}^n$ 
satisfying $|s|=|s'|$ and $|s\oplus s'|=2$, the goal is to identify $S=\{s,s'\}$  by using as few  queries to the oracle as possible. In the following, we will omit  the superscript $S$ in $O^S_{max}$ and  $O^S_{sub}$ for this well-defined \textbf{HSP}.\\

Since \textbf{HSP} for $n=2$ is trivial such that $S=\{01,10\}=\{10,01\}$, in this paper we focus on the non-trivial problem with $n\geq 3$. As mentioned before, the string learning problem has attracted much attention in both classical and quantum computing,  and we hope  to extend this widely 
studied problem for  developing more applications, especially for solving some problems in matroid theory. Later, we will reveal the link between \textbf{HSP} and matroid theory with detailed discussions in Section \ref{sec:app}.

\subsection{Our Contributions}
%To explore the advantages of quantum mechanics in solving string problems, we investigate string learning problem, not learning a string, but learning a  pair of strings which are almost the same and apply it to the problems of finding the bases of a matroid.
We will show that quantum computing can speed up the solution of \textbf{HSP} by constructing  two  quantum algorithms with query complexities  lower than the classical lower bounds. 
More specifically, we have the following theorems.

\begin{theorem}
  \label{th:rank}
  There is a quantum algorithm using $O(1)$ queries to the max inner product oracle to solve \textbf{HSP}
  with high probability\footnote{In this paper,
  "with high probability" means  with any constant probability greater than 1/2.}.
\end{theorem}

\begin{theorem}
  \label{th:independence}
  There is a quantum algorithm using $\frac{1}{2}n+O(\sqrt{n})$ queries to the sub-set
  oracle to solve \textbf{HSP} with high probability.
  %where $c$ is a constant number independent of $n$.
\end{theorem}

\begin{theorem}
  \label{th:lowerbounds}
 Any classical algorithm for solving \textbf{HSP} requires 
  $\Omega(n/\log_{2}n)$  queries to the max inner product oracle  or  $n+\Omega(1)$ queries to the
  sub-set oracle.
\end{theorem}

The rest of the paper is organised as follows.
%In Section \ref{sec:pre}, we give some basic concepts used quantum computing. 
In Section \ref{sec:qa-rank}, we give a quantum algorithm with the max inner product  oracle to prove Theorem 
\ref{th:rank}. 
In Section \ref{sec:qa-independence}, we give a quantum algorithm with the sub-set oracle to prove 
Theorem \ref{th:independence}.
In Section \ref{sec:clb}, we prove Theorem \ref{th:lowerbounds} with information theory.
In Section \ref{sec:app}, we present an  application of our theorems about \textbf{HSP} to matroids. 
Finally, in Section \ref{sec:con} we raise some conclusions.

\section{Quantum Algorithms with the Max Inner Product Oracle}\label{sec:qa-rank}
In this section, we will prove Theorem \ref{th:rank} in two steps. First,  we propose a procedure named Algorithm~\ref{al:rank} that can extract useful information about the hidden strings $\{s,s'\}$ (see Lemma \ref{le:rank}). Next,  based on this subroutine,  we further propose Algorithm~\ref{al:hidden-with-rank} to identify  the hidden $\{s,s'\}$  with high probability. 

\subsection{ Information Extraction with the Max Inner Product Oracle}
To identify $\{s,s'\}$ by using as few queries to the max inner product oracle as possible, we need to extract 
as much information as possible with each query.  In the quantum case, we reveal that one query to the max inner product can obtain certain effective information about  $\{s,s'\}$ 
 as Lemma~\ref{le:rank}.

\begin{lemma}
  \label{le:rank}
  There is a quantum algorithm using one query to the max inner product oracle $O_{max}$ for the hidden strings 
  $\{s,s'\}$ in \textbf{HSP}  to obtain each one of the four results $\{s,s',s\wedge s' ,s\vee s'\}$ with 
  probability  $\frac{1}{4}$. 
\end{lemma}

\begin{algorithm}
  \caption{Extracting  useful information with the max inner product oracle.}\label{al:rank}
  \begin{algorithmic}[1]
    \REQUIRE The quantum max inner product oracle $O_{max}$ in Eq.~\eqref{quantum_Or} with $r(x)$ defined  on $s$ and $s'$ in Eq.~\eqref{def:r_x}.
    
    \ENSURE A string in $\{s,s',s\wedge s',s\vee s'\}$.
    \STATE Initialize the $n+m$ qubits to $|0^n\rangle|0^{m-1}1\rangle$, with  $m=\left\lceil {\log _2}(n) \right\rceil$.
    \STATE Apply the unitary transformation $H^{\otimes(n+m)} $.
    \STATE  Apply  the quantum oracle $O_{max}$.
    \STATE Apply the unitary transformation $H^{\otimes (n+m)}$.
    \STATE Measure the first $n$ qubits and  obtain a string  $\tau\in \{s,s',s\wedge s',s\vee s'\}$. 
    \RETURN $\tau$.
  \end{algorithmic}
\end{algorithm}

\begin{proof}[\textbf{Proof.}]
For simplicity,  
we denote the value of $O_{max}(x)$ for \textbf{HSP} by $r(x)$ as
\begin{equation}
\label{def:r_x}
r(x)=\max\{x\cdot s,x\cdot s'\} \ \text{s.t.} \ |s|=|s'|, |s\oplus s'|=2
\end{equation}
for $x\in \{0,1\}^n$ such that $r(x)\in [0,n-1]$, and we also use $O_{max}$ 
as a quantum oracle that acts on the computational basis as:
\begin{equation}\label{quantum_Or}\left| x \right\rangle \left| y \right\rangle \xrightarrow{{{O}_{max}}}\left| x \right\rangle \left| y+r(x) \mod{2^m} \right\rangle,
\end{equation}
where the first  register $\left| x \right\rangle$ is an $n$-qubit query register, and the second register $\left| y \right\rangle$ is the answer register constaining  $m=\lceil {\log _2}(n) \rceil$ qubits to encode a  decimal integer $y$. Based on Eq.~\eqref{quantum_Or}, we can propose  Algorithm~\ref{al:rank} to satisfy  Lemma~\ref{le:rank}, which evolves as: 
\begin{align}
  &|0^n\rangle|0^{m-1}1\rangle \nonumber\\
  \xrightarrow[]{H^{\otimes (n+m)}} &
  \frac{1}{\sqrt{2^n}}\sum_{x\in\{0,1\}^n}|x\rangle
  \frac{1}{\sqrt{2^m}}\sum_{y=0}^{2^m-1}(-1)^{y}|y\rangle \nonumber\\
  \xrightarrow[]{\hspace{0.4cm}O_{max}\hspace{0.4cm}} &
  \frac{1}{\sqrt{2^n}}\sum_{x\in\{0,1\}^n}|x\rangle
  \frac{1}{\sqrt{2^m}}\sum_{y=0}^{2^m-1}(-1)^{y}|y+r(x)\mod{2^m}\rangle \nonumber\\  
  =&
  \frac{1}{\sqrt{2^n}}\sum_{x\in\{0,1\}^n}(-1)^{-r(x)}|x\rangle
  \frac{1}{\sqrt{2^m}}\sum_{y=0}^{2^m-1}(-1)^{y+r(x)}|y+r(x)\mod{2^m}\rangle \nonumber\\  
  =&
  \frac{1}{\sqrt{2^n}}\sum_{x\in\{0,1\}^n}(-1)^{r(x)}|x\rangle
  \frac{1}{\sqrt{2^m}}\sum_{y'=0}^{2^m-1}(-1)^{y'}|y'\rangle \nonumber\\ 
  \xrightarrow[]{H^{\otimes (n+m)} } &
  \sum_{\tau\in\{0,1\}^n}
  \Big(\frac{1}{2^n}\sum_{x\in\{0,1\}^n}(-1)^{r(x)+\tau\cdot x}\Big)|\tau\rangle|0^{m-1}1\rangle. \label{process_algo1}
\end{align}
More generally, the process of Eq.~\eqref{process_algo1}  tells us how to extract  information about a target oracle (e.g.  $r(x)$) in the coefficient of each basis state $|\tau\rangle$ of the output state by querying the oracle only  once. Note the  well-known Bernstein-Vazirani algorithm \cite{DBLP:journals/siamcomp/BernsteinV97}   and its extended version \cite{DBLP:journals/qip/HunzikerM02} have used similar techniques to identify a target hidden string. In our case here, we give some technical treatment for calculating the probability of a measurement outcome.

Without loss of  generality, we assume that $s$ and $s'$ are different in
the $i$-th and $j$-th bits with $i<j$. For all $n$-bit strings $x\in \{0,1\}^n$, we divide them into four categories as
$X_{00}$,  $X_{01}$, $X_{10}$ and $X_{11}$,
according to the values of the $i$-th and $j$-th bits. That is, any  string  
$x\in X_{k_1 k_2}$ for $k_1 k_2\in \{0,1\}^2$ has $x_i=k_1$ and  $x_j=k_2$, leading to $|X_{00}|=|X_{10}|=|X_{01}|=|X_{11}|=2^{n-2}$.
Note the string  $s_0=s\wedge  s'$ belongs to $X_{00}$ and differs from $s$ or $s'$ in one bit, and we have 
\begin{equation}\label{r_x}
  r(x)=\begin{cases}
     s_0\cdot x, & x\in X_{00},\\
       s_0\cdot x+1, & otherwise, 
    \end{cases}
\end{equation} 
by considering the definition of $r(x)$ in Eq.~\eqref{def:r_x}. 
 Based on Eq.~\eqref{r_x}, we can give the probability of getting result $\tau=s$ upon  measuring the ouput state in  Eq.~\eqref{process_algo1} as 

\begin{align*}
  \Pr\big(\tau=s\big) = & \Big|\frac{1}{2^n}\sum_{x\in\{0,1\}^n}(-1)^{r(x)+s\cdot x}\Big|^2\\
  = & \Big|\frac{1}{2^n}\Big(\sum_{x\in X_{00}}(-1)^{s_0\cdot x+s\cdot x} +
  \sum_{x\in X_{10}}(-1)^{s_0\cdot x+1+s\cdot x} +\\
  &\sum_{x\in X_{01}}(-1)^{s_0\cdot x+1+s\cdot x} +  
  \sum_{x\in X_{11}}(-1)^{s_0\cdot x+1+s\cdot x}\Big)\Big|^2\\
  = & \Big|\frac{1}{2^n}\Big(\sum_{x\in X_{00}}(-1)^{s_0\cdot x+s_0\cdot x} +
  \sum_{x\in X_{10}}(-1)^{s_0\cdot x+1+s_0\cdot x+1} + \\
  &\sum_{x\in X_{01}}(-1)^{s_0\cdot x+1+s_0\cdot x} +  
  \sum_{x\in X_{11}}(-1)^{s_0\cdot x+1+s_0\cdot x+1}\Big)\Big|^2\\
  = &\Big|\frac{1}{2^n}\Big(\sum_{x\in X_{00}} 1+\sum_{x\in X_{10}}1
  +\sum_{x\in X_{01}}(-1)+\sum_{x\in X_{11}}1
  \Big)\Big|^2\\
  = & \Big|\frac{1}{2^n}\Big(2^{n-2}+2^{n-2}-2^{n-2}+2^{n-2}\Big)\Big|^2\\
  = & \frac{1}{4}.
\end{align*}
%for the case $s\in X_{10}$ and $s'\in X_{01}$,  which also holds for  another case
%$s\in X_{01}$ and  $s'\in X_{10}$.
In a similar way, we can also compute
$\Pr\big(\tau=s'\big)=\Pr\big(\tau=s \wedge   s'\big)=\Pr\big(\tau=s\vee s'\big)=\frac{1}{4}$. 
% These imply that
% $\Pr\big(\tau\notin\{s,s',s\wedge s',s\vee s'\}\big)=0$. \\
\end{proof}

\subsection{Proof of Theorem \ref{th:rank}}
In this section, we prove Theorem~\ref{th:rank} by designing Algorithm~\ref{al:hidden-with-rank} based on previous Algorithm~\ref{al:rank}. For convenience, we restate Theorem \ref{th:rank} here.
\begin{th:rank}
  There is a quantum algorithm using $O(1)$ queries to the  max inner product oracle to solve \textbf{HSP}
  with high probability.
\end{th:rank}

\begin{algorithm}
  \caption{Quantum algorithm for learning hidden strings with the max inner product  oracle.}\label{al:hidden-with-rank}
  \begin{algorithmic}[1]
    \REQUIRE The quantum max inner product oracle $O_{max}$ in Eq.~\eqref{quantum_Or} with $r(x)$ defined  on $s$ and $s'$ in Eq.~\eqref{def:r_x}.
    \ENSURE  $\{s,s'\}$ with success probability 0.96899.
    \STATE $\omega\leftarrow 7$.
    \FOR{$i\leftarrow 1$ to $\omega$}
    \STATE Run Algorithm \ref{al:rank} and obtain a string denoted $s_i$.
    \ENDFOR
    \IF{$\exists\;i\neq j$ such that $s_i\neq s_j$ \emph{and}  $|s_i| = |s_j|$}
    \STATE $s\leftarrow s_i,s'\leftarrow s_j$.
    \ELSIF{$\exists\;i,j$ such that $|s_i|-|s_j|=2$ }
    \STATE Find two  indices  $l,l'$ such that $s_i[l]\neq s_j[l]$ and $s_i[l']\neq s_j[l']$.
    \STATE $s\leftarrow s_j\vee{e_l},s'\leftarrow s_j\vee{e_{l'}}$.
    \ELSE\RETURN NULL.
    \ENDIF    
    \RETURN $\{s,s'\}$.
  \end{algorithmic}
\end{algorithm}

\begin{proof}[\textbf{Proof.}]
  We will show that Algorithm \ref{al:hidden-with-rank} uses $7 $ queries to the max inner product
  oracle $O_{max}$ in Eq.~\eqref{quantum_Or} to obtain the hidden string pair $\{s,s'\}$ in Eq.~\eqref{def:r_x} with a success  probability higher than
  $0.95$.

When we run  Algorithm \ref{al:rank} $\omega$  
$(\omega\geq 2)$ times to get $\omega$ string results denoted $T=\{s_1,\dots,s_{\omega}\}$, 
there are two cases that we can definitely  identify $\{s,s'\}$:  
\begin{itemize}
\item[(\textbf{C1})] If there exist  two distinct $s_i$ and $s_j$ in $T$ such that $|s_i|=|s_j|$, then we can
  immediately  identify the hidden strings  as $s_i$ and $s_j$.
\item[(\textbf{C2})] If there are two distinct $s_i$ and $s_j$ in $T$ such that $|s_i|-|s_j|=2$ with $s_i[l]\neq s_j[l]$ and $s_i[l']\neq s_j[l']$
, it means that the target
$s$ and $s'$ differ in two bits $l$ and $l'$ and $s_j=s \wedge s'$. Therefore, we can identify $\{s,s'\}$ as $ \{s_j\vee{e_l}, s_j\vee{e_{l'}}\}$.

\end{itemize}

We now discuss the success probability of  identifying  $\{s,s'\}$ after  running Algorithm~\ref{al:rank} $\omega$ times. 
The notation $\Pr(a,b,c,d)$ is used to denote
  the probability that  $s$, $s'$, $s\wedge s'$ and $s\vee s'$
appear $a$, $b$, $c$ and $d$ times respectively in the
$\omega$ results $T=\{s_1,\ldots,s_{\omega}\}$.
By \emph{multinomial distribution} and Lemma \ref{le:rank}, we have
\begin{equation}
\label{Pr_abcd}
  \Pr(a,b,c,d)=\frac{\omega!}{a!\cdot b!\cdot c!\cdot d!} (\frac{1}{4})^{\omega}.
\end{equation}
Then the  probability of identifying $\{s,s'\}$ is given by summing up the probabilities of all events satisfying  (\textbf{C1}) or 
(\textbf{C2}) as
\begin{equation}
\label{Pr_identify}
  \Pr(\text{Identify}\{s,s'\}) = \sum_{\textbf{(C1) or (C2)}}\Pr(a,b,c,d),
\end{equation}
where the case  \textbf{(C1)} or \textbf{(C2)} corresponds to:
\begin{equation}
\label{C1_C2}
  \left\{ 
  \begin{array}{r}
    a,b,c,d\;\;\text{are non-negative integers};\\
    a+b+c+d=\omega;\\
    ab>0 \;\text{or}\; cd>0.\\
  \end{array}
  \right.
\end{equation}

By combining Eqs.~\eqref{Pr_abcd},\eqref{Pr_identify} and \eqref{C1_C2}, we can solve the inequality  $ \Pr(\text{Identify}\{s,s'\})$ $\geq 0.95 $ and obtain
$\omega\geq 7$. As a result, the above analysis with a  parameter $\omega=7$ is formulated as Algorithm~\ref{al:hidden-with-rank}, where we can identify
$\{s,s'\}$ of any size $n$  with a  success probability $0.96899$. Thus, we prove Theorem~\ref{th:rank} such that a constant number of queries  to the max inner
product oracle can solve \textbf{HSP} with high probability.
\end{proof}

\section{Quantum Algorithms with the Sub-set Oracle}\label{sec:qa-independence}
In this section, we will prove Theorem \ref{th:independence} in two steps. First we use the 
technique proposed by Dam \cite{DBLP:conf/focs/Dam98} to give a quantum
algorithm based on  the sub-set oracle  to output $s\vee s'$ with high probability  (see Lemma \ref{le:independence}).
Then we use Grover's algorithm to find $s$ and $s'$ in the set consisting of the
largest proper subset of  $s\vee s'$. 

\subsection{ Information Extraction with the Sub-set Oracle}
Inspired by Ref. \cite{DBLP:conf/focs/Dam98},  we can use
$\frac{n}{2}+O(\sqrt{n})$ queries to the general  sub-set oracle $O_{sub}$ in
Eq.~\eqref{O_sub_S} to learn $t=\bigvee_{s\in S} s$, i.e., the bitwise \textbf{OR}
of the strings in $S$ with arbitrarily small error probability, which would be
useful for further  identifying the case of $S=\{s,s'\}$.

\begin{lemma}
  \label{le:independence}
  There is a quantum algorithm using $\frac{n}{2}+O(\sqrt{n})$ queries
  to $O_{sub}$ in Eq.~\eqref{O_sub_S}  to learn $t=\bigvee_{s\in S} s$, i.e.,
  the bitwise \textbf{OR} of the strings in $S$,  with an arbitrarily
  small error probability.
  %more than $0.95$.
\end{lemma}

\begin{algorithm}
  \caption{Extracting useful information with the sub-set oracle.}\label{al:independence}
  \begin{algorithmic}[1]
    \REQUIRE A sub-set oracle $O_{sub}$  in Eq.~\eqref{O_sub_S}; a parameter $k=n/2+O(\sqrt{n})$.
    \ENSURE The bitwise \textbf{OR} of  strings in $S$ with arbitrarily small
    probability.
    \STATE Initialize the $(n+1)$ registers to $|0^n\rangle|1\rangle$.
    \STATE Apply the unitary transformation $U_k\otimes H$ with $U_k$ in Eq.~\eqref{eq:uk}.
    \STATE Apply the  unitary transformation $A_k$ in Eq.~\eqref{eq:ak}.
    \STATE Apply the unitary transformation $H^{\otimes n}\otimes H$.
    \STATE Measure the first $n$-qubit  register and  obtain $\tau\in\{0,1\}^n$.
    \RETURN $\tau$.
  \end{algorithmic}
\end{algorithm}

\begin{proof}[\textbf{Proof.}]
  We will show that Algorithm \ref{al:independence} can obtain the bitwise \textbf{OR} of the strings in $S$ with arbitrarily small error probability
  by comsuming $\frac{n}{2}+O(\sqrt{n})$ 
  queries to $O_{sub}$. Let $t$ be the bitwise \textbf{OR} of strings in $S$.
  
  At first, we introduce how to  obtain the inner product of  $t$ and $x\in \{0,1\}^n$ by querying the oracle $O_{sub}$.   
 It is easy to see that for any $j$, $O_{sub}(e_j)=1$ if and only if $t_j=1$. Thus, we have
\begin{equation}
\label{bitwise_x_t}
x\cdot t = \sum_{j\in [n]}x_j t_j = \sum_{j\in [n]}x_j O_{sub}(e_j) ,
\end{equation}
which indicates that we can use $|x|$ queries to $O_{sub}$ to  calculate the value
 $x\cdot t $ for any $x\in \{0,1\}^n$ and $t=\bigvee_{s\in S} s$.
  
   Next, we construct an unitary transformation $A_k$ for a given threshold number $k$ such that 
\begin{equation}\label{eq:ak}
  A_{k}|x\rangle|b\rangle =
  \left\{
      \begin{array}{ll}
        |x\rangle |b\oplus (x\cdot t \mod{2})\rangle & \text{if}\;|x|\leq k,\\
        |x\rangle|b\rangle & \text{if}\;|x|> k\\        
      \end{array}
  \right.
\end{equation}  
 as shown in Figure \ref{fig:Ak}, which consists of a series of  operators as:
 \begin{figure}
    \centering
    \includegraphics[width=0.9\textwidth,height=0.6\textwidth]{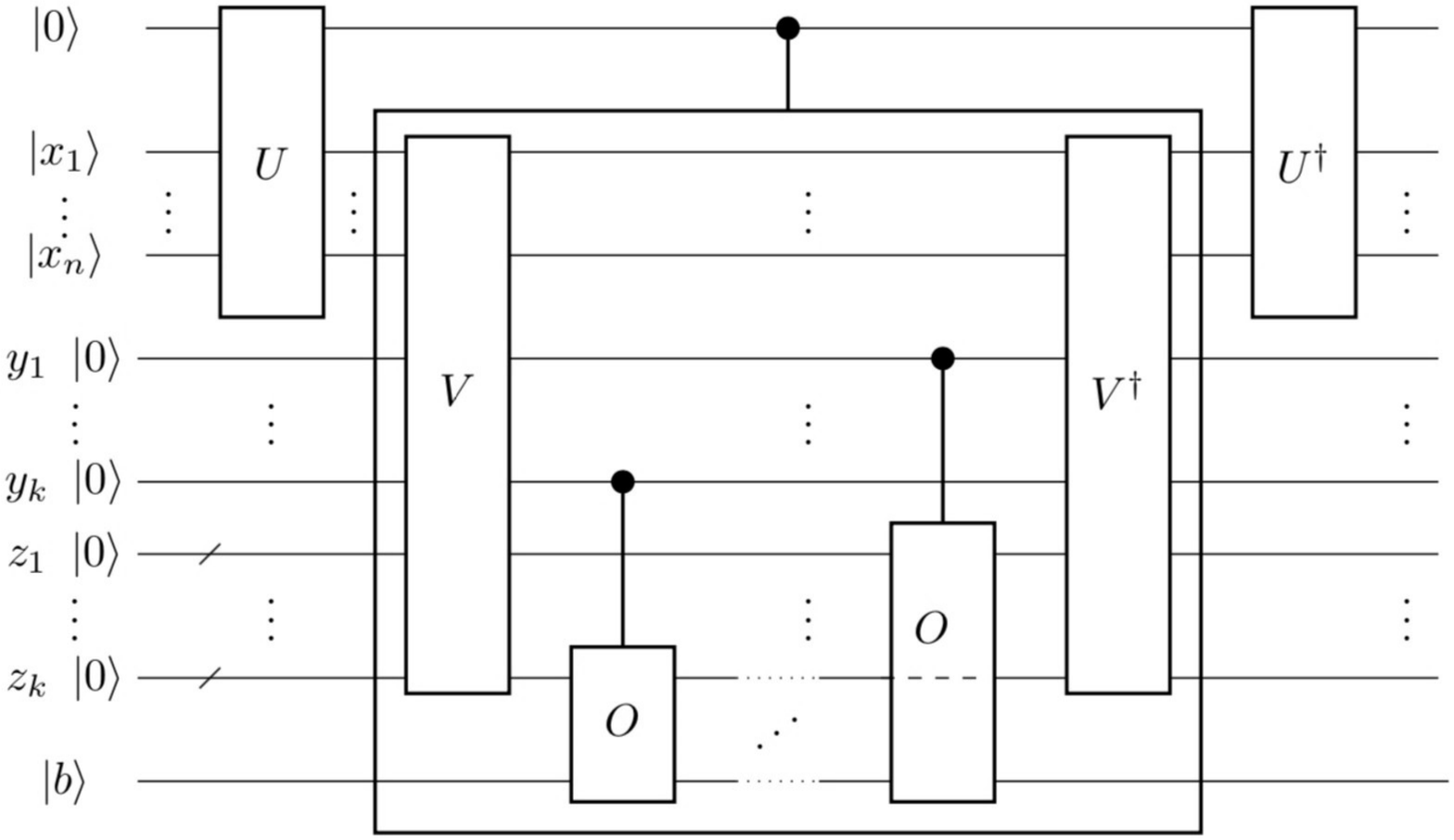}
    \caption{An implementation of $A_k$ with $k$ sub-set oracle $O_{sub}$.}
\label{fig:Ak}
\end{figure}
 
\begin{enumerate}
  \item[(1)] The unitary operation $U$ acts on a qubit state $|a\rangle$ and an $n$-qubit register $|x\rangle=|x_1\rangle \ldots |x_n\rangle $, and tests  whether the Hamming weight $|x|$ is less than or equal to $k$, that is
\begin{equation}
\label{def:U_operator}
U|a\rangle |x\rangle=|a\oplus h_{k}(x)\rangle |x\rangle,
\end{equation}
where $h_{k}(x)=1$ if
$|x|\leq k$, otherwise $h_{k}(x)=0$. For  $|x|\leq k$, the boxed module in the middle
of Figure \ref{fig:Ak} is executed.   
  \item[(2)]\label{V_operator}  The unitary operation $V$ copies up to the first $k$ 1s in $x=x_1\cdots x_n$ to 
$y_1$ to $y_k$, and store the corresponding position where 1 appears in $z_1$ to $z_k$ in turn.
%  The unitary operation $V$ acts on the $n$-qubit $|x\rangle=|x_1\rangle \ldots |x_n\rangle $, $k$-qubit $|y\rangle=|y_1\rangle \ldots |y_k\rangle $, and $k$ $m$-qubit registers denoted $z_i(i=1,2,\ldots,k)$ with $m=\lceil {\log_2{n}} \rceil$. Denote the index set of 1s in $x$ as $\{j_1,j_2,\ldots j_{|x|}\}$ such that $x_{j_i}=1$ for $i\in[1,|x|]$, and the operator $V$ has the effect
%\begin{align}
%V&|x\rangle |y=0^{k}\rangle (|z_1=0^{m}\rangle... |z_k=0^{m}\rangle) \nonumber\\ = & \begin{cases}|x\rangle |y=1^{|x|}0^{k-|x|}\rangle (|j_1\rangle...|j_{|x|}\rangle |z_{|x|+1}=0^{m}\rangle... |z_k=0^{m}\rangle), & |x|\leq k;\\ |x\rangle |y=0^{k}\rangle (|z_1=0^{m}\rangle... |z_k=0^{m}\rangle), &|x|>k,
%\end{cases}
%\end{align}  
%where each  $|j_i\rangle$  with $i\in [1,|x|]$ is  represented by the $m$-qubit register $|z_i\rangle$. 

  \item[(3)] The  unitary operation controlled-$O$ acts on the control qubit  $|y_i\rangle$ and target qubits in  $|z_i\rangle$ and $|b\rangle$ with $i=1,2,\ldots,k$, such that $O$ has the effect
\begin{equation}
\label{def: O_operator}   
{O}|{z_i}\rangle|b\rangle=|{z_i}\rangle|b\oplus O_{sub}(e_{z_i})\rangle.
\end{equation}

  %\item[(4)] The unitary operators $V^\dagger$ and $U^\dagger$.

\end{enumerate}

%As shown in Figure~\ref{fig:Ak}, it can bee derived that the above operators together can  implement the transformation
%\begin{align}
%     &|a=0\rangle |x\rangle |y=0^{k}\rangle (|z_1=0^{m}\rangle... |z_k=0^{m}\rangle)|b\rangle \nonumber \\ \rightarrow &  \begin{cases}
 %   |a=0\rangle |x\rangle |y=0^{k}\rangle (|z_1=0^{m}\rangle... |z_k=0^{m}\rangle)|b\oplus (x\cdot t \mod{2})\rangle, & |x|\leq k ;\\
 %      |a=0\rangle |x\rangle |y=0^{k}\rangle (|z_1=0^{m}\rangle... |z_k=0^{m}\rangle)|b\rangle, & |x|> k 
 %   \end{cases}  
%\end{align}
%and thus realize $A_k$ in Eq.~\eqref{eq:ak} by regarding $|a \rangle $,$|y \rangle$ and $|z_1 \rangle ...|z_k \rangle$ as ancilla registers, which would consume $k$ sub-set oracles $O_{sub}^S$.
%\input{pic-Ak}

Finally, for $k$ we define an integer
\begin{equation}
\label{def:M_k}
M_{k}=\sum_{i=0}^{k}\binom{n}{i}
\end{equation}
 and an $n$-qubit unitary transformation  $U_{k}$ 
independent of   $O_{sub}$ such that
  \begin{equation}\label{eq:uk}
    U_{k}|0\rangle = \frac{1}{\sqrt{M_{k}}}\sum_{x\in\{0,1\}^n:|x|\leq k}|x\rangle.
  \end{equation}

Based on the introduced  Eqs.~\eqref{eq:ak} and~\eqref{eq:uk}, the Algorithm~\ref{al:independence} evolves as:

  \begin{align*}
   &|0^n\rangle|1\rangle\\
  \xrightarrow[]{U_{k}\otimes H} &
  \frac{1}{\sqrt{M_{k}}}\sum_{x\in\{0,1\}^n:|x|\leq k}\big|x\rangle|-\rangle\\
  \xrightarrow[]{\hspace{0.4cm}A_{k}\hspace{0.4cm}} &
  \frac{1}{\sqrt{M_{k}}}\sum_{x\in\{0,1\}^n:|x|\leq k}(-1)^{x\cdot t}\big|x\rangle|-\rangle\\
  \xrightarrow[]{H^{\otimes n}\otimes H} &
  \sum_{\tau\in\{0,1\}^n}
  \Big(\frac{1}{\sqrt{M_{k}2^{n}}}\sum_{x\in\{0,1\}^n:|x|\leq k}(-1)^{x\cdot t\oplus x\cdot \tau}\Big)\big|\tau\rangle|1\rangle,\\
  \end{align*}
for any value of $k$, and thus the  probability of getting result $\tau=t$ upon measuring the ouput state is 
  \begin{align}
  \label{Pr_Mk}
    \Pr\Big(\tau=t\Big)&=
    \Big|\frac{1}{\sqrt{M_{k}2^n}}\sum_{x\in\{0,1\}^n:|x|\leq k}
    (-1)^{x\cdot t\oplus x\cdot t}\Big|^2 \nonumber\\
    &=\frac{M_k}{2^n}.
  \end{align}

Now we consider the value range of $k$ that enables the probability in Eq.~\eqref{Pr_Mk} to exceed 0.95.
Let $X_1,\dots,X_n$ be $n$ random variables such that  each $X_j (j\in [n])$ represents the value of the $j$th bit 
 in a random string $x\in\{0,1\}^n$. Obviously, $X_1,\dots,X_n$ are
independent and they are the $0-1$ distribution with probability $1/2$. Let
$X=\sum_{i=1}^{n}X_{i}$. By a simple calculation 
we have 
\begin{align}
   \Pr(X\leq k)={M_k}/{2^n} =\Pr({\tau=t}).
\end{align}
Therefore, we can get $\Pr({\tau=t})$ by  estimating $\Pr(X\leq k)$. Also note that the expectation and variance of $X$ are
$\mathbb{E}[X]=\frac{1}{2}n$ and $D(X)=\frac{1}{4}n$,   respectively. 
By De Moiver-Laplace Theorem we have 
  \begin{equation}
  \label{limPr}
   \lim_{n\rightarrow \infty}\Pr\Big(X\leq \frac{n}{2}+\frac{q\sqrt{n}}{2}\Big)= \lim_{n\rightarrow \infty}\Pr\Big(\frac{X-\mathbb{E}[X])}{\sqrt{D(X)}}\leq q\Big) =
    \int_{-\infty}^{q}\frac{1}{\sqrt{2\pi}}e^{-\frac{t^2}{2}}\,dt
  \end{equation}
for any $q\in\mathbb{R}$.
Obviously Eq.~\eqref{limPr} monotonically increases with $q$. By letting Eq.
\eqref{limPr} $>0.95$, we obtain $q> 1.6449$. Thus for quite large $n$, the integer parameter taken as 
$k=\lfloor n/2+\sqrt{n} \rfloor$  in Algorithm~\ref{al:independence}  corresponds to  $q\approx2>1.6449$ in  Eq.~\eqref{limPr},
such that  the probability of obtaining $t=\bigvee_{s\in S} s$ is  $\Pr(\tau=t)=\Pr(X\leq k)>0.95$. 

We also examine how Algorithm~\ref{al:independence} works with small and moderate $n$. Our numerical calculation shows that the
probability curve  $\Pr(\tau=t)$ in Eq.~\eqref{Pr_Mk} oscillates between  $0.9648$ and $1$ for  $3\leq n \leq 1000$ with
$k=\lfloor n/2+\sqrt{n} \rfloor$, and seems to approach the case $q=2$ and  Eq.~\eqref{limPr}=0.9773 as $n$ increases (e.g.
$\Pr(\tau=t)=0.9770,0.9786,0.9769$ for $n=998,999,1000$). 

Actually  the error probability can be made  arbitrarily small by letting  $k=\frac{n}{2}+\lambda\sqrt{n}$, which is explained below. 
The complementary unit Gaussian distribution function is defined as
\begin{equation}\label{qfunc}
    Q(x)=1-\int_{-\infty}^{x}\frac{1}{\sqrt{2\pi}}e^{-\frac{t^2}{2}}\,dt=
    \int_{x}^{\infty}\frac{1}{\sqrt{2\pi}}e^{-\frac{t^2}{2}}\,dt.
\end{equation}
For $x\geq 0$, it is known that 
\begin{equation}\label{qfunc-bound}
    Q(x)\leq \frac{1}{2}e^{-\frac{x^2}{2}}
\end{equation}
with equality holding only at $x=0$. 
Let $k=\frac{n}{2}+\lambda\sqrt{n}$. By Eq.~\eqref{limPr}, Eq.~\eqref{qfunc} and
Eq.~\eqref{qfunc-bound}, for a large $n$, we have
\begin{equation}
    \Pr(X>k) = \Pr(X>\frac{n}{2}+\lambda\sqrt{n}) = Q(2\lambda)\leq \frac{1}{2}
    e^{-2\lambda^2}
\end{equation}
This means that  the error probability   $\Pr({\tau\neq t})$ can be made arbitrarily small with 
$\frac{n}{2}+O(\sqrt{n})$ queries to the sub-set oracle.
\end{proof}
%Besides, we can also obtain $t$ with arbitrarily  high probability by setting 
%$k= n/2+O(\sqrt{n})$ in Algorithm~\ref{al:independence},  corresponding to a certain %value of $q$ in Eq.~\eqref{limPr}. 
%For example, $k=\lfloor n/2+2\sqrt{n} \rfloor$  for $n\geq 12$ (corresponding to $q=4$) %would enable the probability
%$\Pr(\tau=t)$ 
%in Eq.~\eqref{Pr_Mk}  to be higher than 0.9999. 

\subsection{Proof of Theorem \ref{th:independence}}
For convenience, we  restate Theorem \ref{th:independence} here.
\begin{th:ind}
  There is a quantum algorithm using $\frac{1}{2}n+O(\sqrt{n})$ queries to the sub-set
  oracle to solve \textbf{HSP} with high probability.
\end{th:ind}

\begin{proof}[\textbf{Proof.}]
 
  First, by Lemma \ref{le:independence},  we can use   Algorithm \ref{al:independence} to obtain $t=s \vee s'$ with an arbitrarily small 
  error probability, which 
  requires $\frac{1}{2}n+O(\sqrt{n})$ queries to the sub-set oracle.

  Let $\Delta_t$ be the set of strings  obtained by replacing exactly one $1$ in $t$ by $0$. Then there is $|\Delta_t|=|t|-1$. 
  Then we can use Grover's algorithm to find the pair $s,s'$ in $\Delta_t$  satisfying $O_{sub}(s)=1$ and 
  $O_{sub}(s')=1$. This needs $O(\sqrt{n})$ queries to the sub-set oracle.

  Therefore,  Algorithm \ref{al:hidden-with-independence}  consumes
  $\frac{1}{2}n+O(\sqrt{n})+O(\sqrt{n})= \frac{1}{2}n+O(\sqrt{n})$ queries to 
  the sub-set oracle to obtain $\{s,s'\}$ with high probability.
\end{proof}
\begin{algorithm}
  \caption{Quantum algorithm for learning hidden strings with the sub-set oracle.}
  \label{al:hidden-with-independence}
  \begin{algorithmic}[1]
    \REQUIRE The sub-set oracle $O_{sub}$  in Eq.~\eqref{O_sub_S} with the hidden  $S=\{s,s'\}$  satisfying 
    $|s|=|s'|$ and $|s\oplus s'|=2$.
    \ENSURE $\{s,s'\}$ with high probability.
    \STATE $t\leftarrow $ Algorithm \ref{al:independence}.
    \STATE Apply Grover's algorithm to find the pair $s,s'$ in $\Delta_t$  satisfying $O_{sub}(s)=1$ and 
    $O_{sub}(s')=1$.
    \RETURN $\{s,s'\}$.
  \end{algorithmic}
\end{algorithm}

\section{Classical Lower Bounds}
\label{sec:clb}
In this section, we give the  information-theoretic lower bound
on \textbf{HSP} to prove Theorem~\ref{th:lowerbounds}. 
Compared with  Theorem~\ref{th:lowerbounds},  our quantum algorithms show substantial  speedups over  classical  counterparts.  First we restate Theorem \ref{th:lowerbounds} in the following.
\begin{th:low}
   Any classical algorithm for solving \textbf{HSP} requires 
  $\Omega(n/\log_{2}n)$  queries to the max inner product oracle  or  $n+\Omega(1)$ queries to the
  sub-set oracle.
\end{th:low}

\begin{proof}[\textbf{Proof.}]
Let $N$ be the number of all possible solutions $\{s,s'\}$ to \textbf{HSP}  satisfying the promise $|s\oplus s'|=2$ and $|s|=|s'|$. Then there is
\begin{equation}
\label{total_N}
  N =\binom{n}{2}\cdot 2^{n-2}=n(n-1)\cdot 2^{n-3}.
\end{equation}
By noting that $\{s,s'\}$ and $\{s',s\}$ are the same string pair.
Considering the range of oracle functions $O_{max}(x)$ and  $O_{sub}(x)$ defined in Eq.~\eqref{O_max_S} and Eq.~\eqref{O_sub_S} restricted to $S=\{s,s'\}$,  respectively, a deterministic classical algorithm for  \textbf{HSP}
 with the max inner product oracle or the sub-set oracle can be described as an $n$-ary or binary tree with  $N$ leaf nodes. Therefore, the query complexity of a  classical algorithm is the height of the tree, and the lower bound for the query complexity of \textbf{HSP} is the minimum height of a tree with $N$ leaf nodes.

Let $C_{i}(n)$ and $C_{s}(n)$ be the 
 lower bound of \textbf{HSP} with the max inner product oracle and the sub-set  oracle, 
respectively. Then we can derive that 
\begin{align}
  C_{i}(n)&=\lceil \log_{n}(N) \rceil=\lceil \frac{\log_{2}(n(n-1)2^{n-3})}{\log_{2}(n)} \rceil=\Omega(n/\log_2(n)),\\
  C_{s}(n)&=\lceil \log_{2}(N)  \rceil=\lceil n-3+\log_{2}(n(n-1)) \rceil=n+\Omega(1).
\end{align}
%and for brevity we say that any classical algorithm needs $n+\Omega(1)$  queries to the sub-string oracle to identify $\{s,s'\}$.

\end{proof}

\section{Application to Matroids}\label{sec:app}
With the rapid development of quantum computing, finding more problems that can take 
advantage of quantum speedup has become one of the focus issues in the field of quantum 
computing. 
Solving matroid problems is a potential application of quantum computing.
Huang et al.\cite{huang2021quantum} showed that quadratic quantum speedup is possible
for the calculation problem of finding the girth or the number of circuits (bases, 
flats, hyperplanes) of a matroid, and for the decision problem of deciding whether a matroid is uniform , Eulerian or paving.

The  problems related to bases of a  matroid is very important in matroid theory,
including finding a base (with maximum/minimum weigh), enumerating all the bases, counting
the number of base and so on. These problems are of practical significance, 
for example, finding a base of a vector space or a (maximum/minimum weigh) 
spanning tree of a graph are the special case of finding a base of a matroid \cite{edmonds1971matroids}.
Counting the number of base has also been extensively studied \cite{feder1992balanced,
azar1994problem,anari2018log,anari2019log,anari2020isotropy}, and determining the reliability of a graph is a special case of it.
Enumerating the base of a matroid is interesting and also extensively studied. There are 
many works on how to efficiently enumerate the bases of a matroid or the spanning trees of 
a graph \cite{DBLP:journals/dam/AvisF96, DBLP:journals/arscom/NeudauerMS03, 
DBLP:journals/siamdm/KhachiyanBEGM05, DBLP:conf/esa/KhachiyanBBEGM06,
DBLP:journals/jct/Maxwell09, DBLP:conf/soda/CardinalMM22,DBLP:conf/fun/MerinoMW22}.
Due to the importance of the base problems and the potential speedup of 
quantum algorithms, we consider the following problem.\\

\noindent\textbf{Identify Matroids' Bases.} Given a $2$-bases matroid which has $2$ bases and can be accessed 
by a  matroid oracle (\emph{independence oracle} or \emph{rank oracle}), how many oracle
queries are required to identify its bases?   \\

In this section, we will show how to transform the above problem  to the hidden string problem, so that the  previous obtained quantum algorithms  can be applied.

\subsection{Matroid}
Here we give some basic definitions and concepts on matroids. 
Matroid theory was established as a generalization of linear algebra and graph
theory. Some concepts are similar to those of linear algebra or graphs.
One can refer to \cite{OUP/oxley11} or \cite{welsh1976matroid} for more details
about matroid theory.

\begin{definition}[\textbf{Matroid}]
  \label{def:matroid}
  A \emph{matroid} is a combinational object defined by the tuple
  $M=(E,\mathcal{I})$ on the finite ground set $E$ and $\mathcal{I}\subseteq 2^E$ such
  that the following properties hold:
  \begin{enumerate}
  \item[\textbf{I0}.] $\emptyset \in \mathcal{I}$;
  \item[\textbf{I1}.] If $A'\subseteq A$ and $A\in\mathcal{I}$, then $A'\in\mathcal{I}$;
  \item[\textbf{I2}.] For any two sets $A,B\in\mathcal{I}$ with $|A|<|B|$, there exists
    an element $x\in B-A$ such that $A\cup\{x\}\in\mathcal{I}$.
  \end{enumerate}
\end{definition}
The members of $\mathcal{I}$ are the \emph{independent} sets of $M$.
A subset of $E$ not belonging to $\mathcal{I}$ is called \emph{dependent}.
A \emph{base} of $M$ is a maximal independent subset of $E$, and the collection of bases
is denoted by $\mathcal{B}(M)$. For a positive integer $k$, we call a matoid $M$ is a $k$-bases 
matroid (on $E$) if $|\mathcal{B}(M)|=k$.

\begin{definition}[\textbf{Rank}]
  \label{def:rank}
  The \emph{rank function} of  matroid $M=(E,\mathcal{I})$ is a function
  $r:2^E\rightarrow\mathbb{Z}$ defined by
  \[
   r(A) = \max\{|X|:X\subseteq A,X\in\mathcal{I}\}\;\;\;\;\;\;(A\subseteq E)
  \]
\end{definition}
The rank of $M$ sometimes denoted by $r(M)$ is $r(E)$.

\begin{definition}[\textbf{Matroid Oracles}]
  \label{def:matroid-oracle}
  Given a matroid $M=(E,\mathcal{I})$, assume we can only access it by querying the
  independence oracle $O_{i}$ or the rank oracle $O_{r}$. For a subset $S\subseteq E$,
  $O_{i}(S)=1$ if $S$ is an independent set of $M$, otherwise $O_{i}(S)=0$ and  
  $O_{r}(S)=r(S)$. 
  \end{definition}
  
\subsection{Identify 2-bases Matroids' Bases}
One way of representing a matrod is to represent its independent set with
a binary indicator vector. Specifically, we first list the elements of $E$
as $\{e_1,\dots,e_n\}$, and we also use $e_j$ to denote an $n$-bit 0-1
string, where the $j$-th bit is 1 and the other bits are 0. Then for any
$A\subseteq E$, we use the $n$-bit string $\bigoplus_{e\in A}e$ as the 
indicator vector of $A$. We define $\chi:2^E\rightarrow\{0,1\}^{n}$ as 
the mapping function of the set to its indicator vector by
\begin{align*}
    \chi(A)=\bigoplus_{e\in A}e \;\;\;\;\;\;(A\subseteq E).
\end{align*}
%And define $\chi^{-1}:\{0,1\}^{n}\rightarrow 2^E$ as the inverse of
%$\chi$ by
%\begin{align*}
%    \chi^{-1}(s) = \{e_j:s[j]=1,j\in[n]\} \;\;\;\;\;\;(s\subseteq %\{0,1\}^n).
%\end{align*}

\noindent\textbf{Max Inner Product Oracle and Rank Oracle.}
From the definition of matroid, we know that for any independent set 
$J\in\mathcal{I}(M)$, there must be some base $B\in\mathcal{B}(M)$ that contains $J$. Then the rank function of matroid $M$ can be restate as
\begin{align*}
    r(A)=\max\{|A\cap B|:B\in\mathcal{B}(M)\}\;\;\;\;\;\;(A\subseteq E).
\end{align*}
If the subsets of $E$ are represented by binary indicator vectors, the cardinality of the intersection of two sets is equal to the inner product of their
corresponding indicator vectors. Thus, the rank function of matroid $M$ can also be described as 
\begin{align*}
    r(A)=\max\{\chi(A)\cdot\chi(B):B\in\mathcal{B}(M)\}\;\;\;\;\;\;(A\subseteq E).
\end{align*}
Comparing the definitions of the \emph{max inner product oracle} and the matroid
\emph{rank oracle}, we can see that they are the same:
\begin{align*}
    O_{r}(A)=r(A)=O_{max}(\chi(A))\;\;\;\;\;\;(A\subseteq E).
\end{align*}

\noindent\textbf{Sub-set Oracle and Independence Oracle.}
By the definition of matroid, we can restate the independence oracle as 
\begin{align*}
     O_i(A)=\left\{
      \begin{array}{ll}
        1\;\;\text{if}\; \exists\;B\in \mathcal{B}(M)\;\text{s.t.}\; A\subseteq B,\\
        0\;\;\text{otherwise}.
      \end{array}
      \right.
\end{align*}
From this definition, we can see that the independence oracle $O_i$ is 
determined by the set of the bases of a matroid. In other words, the independence
oracle $O_i$ hides the bases of a matroid.
Comparing the definitions of  the \emph{sub-set oracle} and the \emph{independence oracle}, we can also see that they are the same.\\

\noindent\textbf{Identify Matroid's Bases and Hidden Strings Problem.} Given a
2-bases matroid $M$,  let $B, B'\in\mathcal{B}(M)$ be its two different bases. By
the definition of matroid (Definition \ref{def:matroid}),  it can be seen that
$|B|=|B'|$ and $|(B-B')\cup (B'-B)|=2$. Correspondingly,   we have $|\chi(B)|=|\chi(B')|$ and $|\chi(B)\oplus\chi(B')|=2$. Therefore, identifying the two bases $B, B'$ is equivalent to
\textbf{HSP} with $S=\{\chi(B), \chi(B')\}$. Therefore, 
Theorem \ref{th:rank} and Theorem \ref{th:independence}  can be applied to  identify the bases of a 2-bases matroid.

\section{Conclusion}
\label{sec:con}
In this paper we consider quantum algorithms for \textbf{HSP} with the max inner product
oracle and the sub-set oracle. We present quantum algorithms using $O(1)$ queries to the max 
inner product oracle and $\frac{1}{2}n+O(\sqrt{n})$ queries to the sub-set oracle to solve 
\textbf{HSP}, respectively. Furthermore,  our quantum algorithms  are applied to identifying the bases of
2-bases matroids.  In the further work, one may consider how to extend \textbf{HSP} to the case with more strings and apply it to more matroid problems or other problems.

\section*{Acknowledgement}
The authors thank Yongzhen Xu and Jingquan Luo for useful discussions.

\bibliography{reference}

\begin{thebibliography}{10}
\expandafter\ifx\csname url\endcsname\relax
  \def\url#1{\texttt{#1}}\fi
\expandafter\ifx\csname urlprefix\endcsname\relax\def\urlprefix{URL }\fi
\expandafter\ifx\csname href\endcsname\relax
  \def\href#1#2{#2} \def\path#1{#1}\fi

\bibitem{DBLP:journals/siamcomp/KnuthMP77}
D.~E. Knuth, J.~H.~M. Jr., V.~R. Pratt,
  \href{https://doi.org/10.1137/0206024}{Fast pattern matching in strings},
  {SIAM} J. Comput. 6~(2) (1977) 323--350.
\newblock \href {https://doi.org/10.1137/0206024} {\path{doi:10.1137/0206024}}.
\newline\urlprefix\url{https://doi.org/10.1137/0206024}

\bibitem{DBLP:journals/jal/Colussi94}
L.~Colussi, \href{https://doi.org/10.1006/jagm.1994.1008}{Fastest pattern
  matching in strings}, J. Algorithms 16~(2) (1994) 163--189.
\newblock \href {https://doi.org/10.1006/jagm.1994.1008}
  {\path{doi:10.1006/jagm.1994.1008}}.
\newline\urlprefix\url{https://doi.org/10.1006/jagm.1994.1008}

\bibitem{DBLP:journals/cacm/BoyerM77}
R.~S. Boyer, J.~S. Moore, \href{https://doi.org/10.1145/359842.359859}{A fast
  string searching algorithm}, Commun. {ACM} 20~(10) (1977) 762--772.
\newblock \href {https://doi.org/10.1145/359842.359859}
  {\path{doi:10.1145/359842.359859}}.
\newline\urlprefix\url{https://doi.org/10.1145/359842.359859}

\bibitem{DBLP:journals/ipl/Moller-NielsenS84}
P.~M{\o}ller{-}Nielsen, J.~Staunstrup,
  \href{https://doi.org/10.1016/0020-0190(84)90015-2}{Experiments with a fast
  string searching algorithm}, Inf. Process. Lett. 18~(3) (1984) 129--135.
\newblock \href {https://doi.org/10.1016/0020-0190(84)90015-2}
  {\path{doi:10.1016/0020-0190(84)90015-2}}.
\newline\urlprefix\url{https://doi.org/10.1016/0020-0190(84)90015-2}

\bibitem{DBLP:conf/focs/MargaritisS95}
D.~Margaritis, S.~Skiena,
  \href{https://doi.org/10.1109/SFCS.1995.492591}{Reconstructing strings from
  substrings in rounds}, in: 36th Annual Symposium on Foundations of Computer
  Science, Milwaukee, Wisconsin, USA, 23-25 October 1995, {IEEE} Computer
  Society, 1995, pp. 613--620.
\newblock \href {https://doi.org/10.1109/SFCS.1995.492591}
  {\path{doi:10.1109/SFCS.1995.492591}}.
\newline\urlprefix\url{https://doi.org/10.1109/SFCS.1995.492591}

\bibitem{DBLP:journals/jcb/SkienaS95}
S.~Skiena, G.~Sundaram,
  \href{https://doi.org/10.1089/cmb.1995.2.333}{Reconstructing strings from
  substrings}, J. Comput. Biol. 2~(2) (1995) 333--353.
\newblock \href {https://doi.org/10.1089/cmb.1995.2.333}
  {\path{doi:10.1089/cmb.1995.2.333}}.
\newline\urlprefix\url{https://doi.org/10.1089/cmb.1995.2.333}

\bibitem{manin1980computable}
Y.~Manin, Computable and uncomputable, Sovetskoye Radio, Moscow 128 (1980).

\bibitem{IJOTP/Feynman82}
R.~P. Feynman, \href{https://doi.org/10.1007/BF02650179}{Simulating physics
  with computers}, International Journal of Theoretical Physics 21~(6) (1982)
  467--488.
\newblock \href {https://doi.org/10.1007/BF02650179}
  {\path{doi:10.1007/BF02650179}}.
\newline\urlprefix\url{https://doi.org/10.1007/BF02650179}

\bibitem{RSPA/Deutsch85}
D.~Deutsch, \href{https://doi.org/10.1098/rspa.1985.0070}{Quantum theory, the
  church-turing principle and the universal quantum computer}, Proceedings of
  the Royal Society of London. Series A, Mathematical and Physical Sciences
  400~(1818) (1985) 97--117.
\newblock \href {https://doi.org/10.1098/rspa.1985.0070}
  {\path{doi:10.1098/rspa.1985.0070}}.
\newline\urlprefix\url{https://doi.org/10.1098/rspa.1985.0070}

\bibitem{RSPA/Deutsch92}
D.~Deutsch, R.~Jozsa, \href{https://doi.org/10.1098/rspa.1992.0167}{Rapid
  solution of problems by quantum computation}, Proceedings of the Royal
  Society of London Series A 439~(1907) (1992) 553--558.
\newblock \href {https://doi.org/10.1098/rspa.1992.0167}
  {\path{doi:10.1098/rspa.1992.0167}}.
\newline\urlprefix\url{https://doi.org/10.1098/rspa.1992.0167}

\bibitem{DBLP:conf/focs/Simon94}
D.~R. Simon, \href{https://doi.org/10.1109/SFCS.1994.365701}{On the power of
  quantum computation}, in: 35th Annual Symposium on Foundations of Computer
  Science, Santa Fe, New Mexico, USA, 20-22 November 1994, {IEEE} Computer
  Society, 1994, pp. 116--123.
\newblock \href {https://doi.org/10.1109/SFCS.1994.365701}
  {\path{doi:10.1109/SFCS.1994.365701}}.
\newline\urlprefix\url{https://doi.org/10.1109/SFCS.1994.365701}

\bibitem{DBLP:journals/siamcomp/BernsteinV97}
E.~Bernstein, U.~V. Vazirani,
  \href{https://doi.org/10.1137/S0097539796300921}{Quantum complexity theory},
  {SIAM} J. Comput. 26~(5) (1997) 1411--1473.
\newblock \href {https://doi.org/10.1137/S0097539796300921}
  {\path{doi:10.1137/S0097539796300921}}.
\newline\urlprefix\url{https://doi.org/10.1137/S0097539796300921}

\bibitem{DBLP:conf/focs/Shor94}
P.~W. Shor, \href{https://doi.org/10.1109/SFCS.1994.365700}{Algorithms for
  quantum computation: Discrete logarithms and factoring}, in: 35th Annual
  Symposium on Foundations of Computer Science, Santa Fe, New Mexico, USA,
  20-22 November 1994, {IEEE} Computer Society, 1994, pp. 124--134.
\newblock \href {https://doi.org/10.1109/SFCS.1994.365700}
  {\path{doi:10.1109/SFCS.1994.365700}}.
\newline\urlprefix\url{https://doi.org/10.1109/SFCS.1994.365700}

\bibitem{DBLP:conf/stoc/Grover96}
L.~K. Grover, \href{https://doi.org/10.1145/237814.237866}{A fast quantum
  mechanical algorithm for database search}, in: G.~L. Miller (Ed.),
  Proceedings of the Twenty-Eighth Annual {ACM} Symposium on the Theory of
  Computing, Philadelphia, Pennsylvania, USA, May 22-24, 1996, {ACM}, 1996, pp.
  212--219.
\newblock \href {https://doi.org/10.1145/237814.237866}
  {\path{doi:10.1145/237814.237866}}.
\newline\urlprefix\url{https://doi.org/10.1145/237814.237866}

\bibitem{Zhang2022}
S.~Zhang, L.~Li, \href{https://doi.org/10.1007/s42514-022-00090-3}{A brief
  introduction to quantum algorithms}, CCF Transactions on High Performance
  Computing 4~(1) (2022) 53--62.
\newblock \href {https://doi.org/10.1007/s42514-022-00090-3}
  {\path{doi:10.1007/s42514-022-00090-3}}.
\newline\urlprefix\url{https://doi.org/10.1007/s42514-022-00090-3}

\bibitem{DBLP:journals/jda/HariharanV03}
R.~Hariharan, V.~Vinay,
  \href{https://doi.org/10.1016/S1570-8667(03)00010-8}{String matching in
  {\~{o}}(sqrt(n)+sqrt(m)) quantum time}, J. Discrete Algorithms 1~(1) (2003)
  103--110.
\newblock \href {https://doi.org/10.1016/S1570-8667(03)00010-8}
  {\path{doi:10.1016/S1570-8667(03)00010-8}}.
\newline\urlprefix\url{https://doi.org/10.1016/S1570-8667(03)00010-8}

\bibitem{DBLP:journals/algorithmica/Montanaro17}
A.~Montanaro, \href{https://doi.org/10.1007/s00453-015-0060-4}{Quantum pattern
  matching fast on average}, Algorithmica 77~(1) (2017) 16--39.
\newblock \href {https://doi.org/10.1007/s00453-015-0060-4}
  {\path{doi:10.1007/s00453-015-0060-4}}.
\newline\urlprefix\url{https://doi.org/10.1007/s00453-015-0060-4}

\bibitem{DBLP:conf/innovations/GallS22}
F.~L. Gall, S.~Seddighin,
  \href{https://doi.org/10.4230/LIPIcs.ITCS.2022.97}{Quantum meets fine-grained
  complexity: Sublinear time quantum algorithms for string problems}, in:
  M.~Braverman (Ed.), 13th Innovations in Theoretical Computer Science
  Conference, {ITCS} 2022, January 31 - February 3, 2022, Berkeley, CA, {USA},
  Vol. 215 of LIPIcs, Schloss Dagstuhl - Leibniz-Zentrum f{\"{u}}r Informatik,
  2022, pp. 97:1--97:23.
\newblock \href {https://doi.org/10.4230/LIPIcs.ITCS.2022.97}
  {\path{doi:10.4230/LIPIcs.ITCS.2022.97}}.
\newline\urlprefix\url{https://doi.org/10.4230/LIPIcs.ITCS.2022.97}

\bibitem{DBLP:conf/soda/AkmalJ22}
S.~Akmal, C.~Jin,
  \href{https://doi.org/10.1137/1.9781611977073.109}{Near-optimal quantum
  algorithms for string problems}, in: J.~S. Naor, N.~Buchbinder (Eds.),
  Proceedings of the 2022 {ACM-SIAM} Symposium on Discrete Algorithms, {SODA}
  2022, Virtual Conference / Alexandria, VA, USA, January 9 - 12, 2022, {SIAM},
  2022, pp. 2791--2832.
\newblock \href {https://doi.org/10.1137/1.9781611977073.109}
  {\path{doi:10.1137/1.9781611977073.109}}.
\newline\urlprefix\url{https://doi.org/10.1137/1.9781611977073.109}

\bibitem{DBLP:conf/swat/CleveIGNTTY12}
R.~Cleve, K.~Iwama, F.~L. Gall, H.~Nishimura, S.~Tani, J.~Teruyama,
  S.~Yamashita,
  \href{https://doi.org/10.1007/978-3-642-31155-0\_34}{Reconstructing strings
  from substrings with quantum queries}, in: F.~V. Fomin, P.~Kaski (Eds.),
  Algorithm Theory - {SWAT} 2012 - 13th Scandinavian Symposium and Workshops,
  Helsinki, Finland, July 4-6, 2012. Proceedings, Vol. 7357 of Lecture Notes in
  Computer Science, Springer, 2012, pp. 388--397.
\newblock \href {https://doi.org/10.1007/978-3-642-31155-0\_34}
  {\path{doi:10.1007/978-3-642-31155-0\_34}}.
\newline\urlprefix\url{https://doi.org/10.1007/978-3-642-31155-0\_34}

\bibitem{2013DNA}
A.~S. Motahari, G.~Bresler, D.~N.~C. Tse, Information theory of {DNA} shotgun
  sequencing, IEEE Transactions on Information Theory 59~(10) (2013)
  6273--6289.

\bibitem{2012mining}
J.~Dhaliwal, S.~J. Puglisi, A.~Turpin, Practical efficient string mining, IEEE
  Transactions on Knowledge and Data Engineering 24~(4) (2012) 735--744.

\bibitem{2022AES}
Z.~Li, B.~Cai, H.~Sun, H.~Liu, L.~Wan, S.~Qin, Q.~Wen, F.~Gao, Novel quantum
  circuit implementation of advanced encryption standard with low costs,
  Science China: Physics, Mechanics and Astronomy 65~(9) (2022) 290311.

\bibitem{ding-zhu/hwang:2000}
D.~Ding-Zhu, F.~K. Hwang, \href{https://doi.org/10.1142/1936}{Combinatorial
  group testing and its applications}, World Scientific, 2000.
\newblock \href {https://doi.org/doi.org/10.1142/1936}
  {\path{doi:doi.org/10.1142/1936}}.
\newline\urlprefix\url{https://doi.org/10.1142/1936}

\bibitem{DBLP:conf/focs/Dam98}
W.~van Dam, \href{https://doi.org/10.1109/SFCS.1998.743486}{Quantum oracle
  interrogation: Getting all information for almost half the price}, in: 39th
  Annual Symposium on Foundations of Computer Science, {FOCS} '98, November
  8-11, 1998, Palo Alto, California, {USA}, {IEEE} Computer Society, 1998, pp.
  362--367.
\newblock \href {https://doi.org/10.1109/SFCS.1998.743486}
  {\path{doi:10.1109/SFCS.1998.743486}}.
\newline\urlprefix\url{https://doi.org/10.1109/SFCS.1998.743486}

\bibitem{DBLP:journals/tcs/IwamaNRT12}
K.~Iwama, H.~Nishimura, R.~Raymond, J.~Teruyama,
  \href{https://doi.org/10.1016/j.tcs.2012.05.039}{Quantum counterfeit coin
  problems}, Theor. Comput. Sci. 456 (2012) 51--64.
\newblock \href {https://doi.org/10.1016/j.tcs.2012.05.039}
  {\path{doi:10.1016/j.tcs.2012.05.039}}.
\newline\urlprefix\url{https://doi.org/10.1016/j.tcs.2012.05.039}

\bibitem{DBLP:journals/qic/AmbainisM14}
A.~Ambainis, A.~Montanaro, \href{https://doi.org/10.26421/QIC14.5-6-4}{Quantum
  algorithms for search with wildcards and combinatorial group testing},
  Quantum Inf. Comput. 14~(5-6) (2014) 439--453.
\newblock \href {https://doi.org/10.26421/QIC14.5-6-4}
  {\path{doi:10.26421/QIC14.5-6-4}}.
\newline\urlprefix\url{https://doi.org/10.26421/QIC14.5-6-4}

\bibitem{xu2022quantum}
Y.~Xu, S.~Zhang, L.~Li, \href{https://arxiv.org/abs/2206.11221v1}{Quantum
  algorithm for learning secret strings and its experimental demonstration},
  arXiv preprint arXiv:2206.11221 (2022).
\newblock \href {https://doi.org/110.48550/arXiv.2206.11221}
  {\path{doi:110.48550/arXiv.2206.11221}}.
\newline\urlprefix\url{https://arxiv.org/abs/2206.11221v1}

\bibitem{li2022winning}
L.~Li, J.~Luo, Y.~Xu, Winning mastermind overwhelmingly on quantum computers,
  arXiv preprint arXiv:2207.09356 (2022).

\bibitem{JHUP/whitney35}
H.~Whitney, \href{http://www.jstor.org/stable/2371182}{On the abstract
  properties of linear dependence}, American Journal of Mathematics 57~(3)
  (1935) 509--533.
\newblock \href {https://doi.org/10.2307/2371182} {\path{doi:10.2307/2371182}}.
\newline\urlprefix\url{http://www.jstor.org/stable/2371182}

\bibitem{DBLP:journals/qip/HunzikerM02}
M.~Hunziker, D.~A. Meyer,
  \href{https://doi.org/10.1023/A\%3A1019868924061}{Quantum algorithms for
  highly structured search problems}, Quantum Inf. Process. 1~(3) (2002)
  145--154.
\newblock \href {https://doi.org/10.1023/A\%3A1019868924061}
  {\path{doi:10.1023/A\%3A1019868924061}}.
\newline\urlprefix\url{https://doi.org/10.1023/A\%3A1019868924061}

\bibitem{huang2021quantum}
X.~Huang, J.~Luo, L.~Li, \href{https://arxiv.org/abs/2111.12900}{Quantum
  speedup and limitations on matroid properties}, arXiv preprint
  arXiv:2111.12900 (2021).
\newline\urlprefix\url{https://arxiv.org/abs/2111.12900}

\bibitem{edmonds1971matroids}
J.~Edmonds, \href{https://doi.org/10.1007/BF01584082}{Matroids and the greedy
  algorithm}, Mathematical programming 1~(1) (1971) 127--136.
\newblock \href {https://doi.org/10.1007/BF01584082}
  {\path{doi:10.1007/BF01584082}}.
\newline\urlprefix\url{https://doi.org/10.1007/BF01584082}

\bibitem{feder1992balanced}
T.~Feder, M.~Mihail, \href{https://doi.org/10.1145/129712.129716}{Balanced
  matroids}, in: Proceedings of the twenty-fourth annual ACM symposium on
  Theory of computing, 1992, pp. 26--38.
\newblock \href {https://doi.org/10.1145/129712.129716}
  {\path{doi:10.1145/129712.129716}}.
\newline\urlprefix\url{https://doi.org/10.1145/129712.129716}

\bibitem{azar1994problem}
Y.~Azar, A.~Z. Broder, A.~M. Frieze,
  \href{https://doi.org/10.1016/0020-0190(94)90037-X}{On the problem of
  approximating the number of bases of a matroid}, Inf. Process. Lett. 50~(1)
  (1994) 9--11.
\newblock \href {https://doi.org/10.1016/0020-0190(94)90037-X}
  {\path{doi:10.1016/0020-0190(94)90037-X}}.
\newline\urlprefix\url{https://doi.org/10.1016/0020-0190(94)90037-X}

\bibitem{anari2018log}
N.~Anari, S.~O. Gharan, C.~Vinzant,
  \href{https://doi.org/10.1109/FOCS.2018.00013}{Log-concave polynomials,
  entropy, and a deterministic approximation algorithm for counting bases of
  matroids}, in: 2018 IEEE 59th Annual Symposium on Foundations of Computer
  Science (FOCS), IEEE, 2018, pp. 35--46.
\newblock \href {https://doi.org/10.1109/FOCS.2018.00013}
  {\path{doi:10.1109/FOCS.2018.00013}}.
\newline\urlprefix\url{https://doi.org/10.1109/FOCS.2018.00013}

\bibitem{anari2019log}
N.~Anari, K.~Liu, S.~O. Gharan, C.~Vinzant,
  \href{https://doi.org/10.1145/3313276.3316385}{Log-concave polynomials ii:
  high-dimensional walks and an fpras for counting bases of a matroid}, in:
  Proceedings of the 51st Annual ACM SIGACT Symposium on Theory of Computing,
  2019, pp. 1--12.
\newblock \href {https://doi.org/10.1145/3313276.3316385}
  {\path{doi:10.1145/3313276.3316385}}.
\newline\urlprefix\url{https://doi.org/10.1145/3313276.3316385}

\bibitem{anari2020isotropy}
N.~Anari, M.~Derezi{\'n}ski,
  \href{https://doi.org/10.1109/FOCS46700.2020.00126}{Isotropy and log-concave
  polynomials: Accelerated sampling and high-precision counting of matroid
  bases}, in: 2020 IEEE 61st Annual Symposium on Foundations of Computer
  Science (FOCS), IEEE, 2020, pp. 1331--1344.
\newblock \href {https://doi.org/10.1109/FOCS46700.2020.00126}
  {\path{doi:10.1109/FOCS46700.2020.00126}}.
\newline\urlprefix\url{https://doi.org/10.1109/FOCS46700.2020.00126}

\bibitem{DBLP:journals/dam/AvisF96}
D.~Avis, K.~Fukuda, \href{https://doi.org/10.1016/0166-218X(95)00026-N}{Reverse
  search for enumeration}, Discret. Appl. Math. 65~(1-3) (1996) 21--46.
\newblock \href {https://doi.org/10.1016/0166-218X(95)00026-N}
  {\path{doi:10.1016/0166-218X(95)00026-N}}.
\newline\urlprefix\url{https://doi.org/10.1016/0166-218X(95)00026-N}

\bibitem{DBLP:journals/arscom/NeudauerMS03}
N.~A. Neudauer, A.~M. Meyers, B.~Stevens, Enumeration of the bases of the
  bicircular matroid on a complete bipartite graph, Ars Comb. 66 (2003).

\bibitem{DBLP:journals/siamdm/KhachiyanBEGM05}
L.~G. Khachiyan, E.~Boros, K.~M. Elbassioni, V.~Gurvich, K.~Makino,
  \href{https://doi.org/10.1137/S0895480103428338}{On the complexity of some
  enumeration problems for matroids}, {SIAM} J. Discret. Math. 19~(4) (2005)
  966--984.
\newblock \href {https://doi.org/10.1137/S0895480103428338}
  {\path{doi:10.1137/S0895480103428338}}.
\newline\urlprefix\url{https://doi.org/10.1137/S0895480103428338}

\bibitem{DBLP:conf/esa/KhachiyanBBEGM06}
L.~Khachiyan, E.~Boros, K.~Borys, K.~M. Elbassioni, V.~Gurvich, K.~Makino,
  \href{https://doi.org/10.1007/11841036\_41}{Enumerating spanning and
  connected subsets in graphs and matroids}, in: Y.~Azar, T.~Erlebach (Eds.),
  Algorithms - {ESA} 2006, 14th Annual European Symposium, Zurich, Switzerland,
  September 11-13, 2006, Proceedings, Vol. 4168 of Lecture Notes in Computer
  Science, Springer, 2006, pp. 444--455.
\newblock \href {https://doi.org/10.1007/11841036\_41}
  {\path{doi:10.1007/11841036\_41}}.
\newline\urlprefix\url{https://doi.org/10.1007/11841036\_41}

\bibitem{DBLP:journals/jct/Maxwell09}
M.~Maxwell, \href{https://doi.org/10.1016/j.jcta.2008.06.007}{Enumerating bases
  of self-dual matroids}, J. Comb. Theory, Ser. {A} 116~(2) (2009) 351--378.
\newblock \href {https://doi.org/10.1016/j.jcta.2008.06.007}
  {\path{doi:10.1016/j.jcta.2008.06.007}}.
\newline\urlprefix\url{https://doi.org/10.1016/j.jcta.2008.06.007}

\bibitem{DBLP:conf/soda/CardinalMM22}
J.~Cardinal, A.~I. Merino, T.~M{\"{u}}tze,
  \href{https://doi.org/10.1137/1.9781611977073.84}{Efficient generation of
  elimination trees and graph associahedra}, in: J.~S. Naor, N.~Buchbinder
  (Eds.), Proceedings of the 2022 {ACM-SIAM} Symposium on Discrete Algorithms,
  {SODA} 2022, Virtual Conference / Alexandria, VA, USA, January 9 - 12, 2022,
  {SIAM}, 2022, pp. 2128--2140.
\newblock \href {https://doi.org/10.1137/1.9781611977073.84}
  {\path{doi:10.1137/1.9781611977073.84}}.
\newline\urlprefix\url{https://doi.org/10.1137/1.9781611977073.84}

\bibitem{DBLP:conf/fun/MerinoMW22}
A.~I. Merino, T.~M{\"{u}}tze, A.~Williams,
  \href{https://doi.org/10.4230/LIPIcs.FUN.2022.22}{All your bases are belong
  to us: Listing all bases of a matroid by greedy exchanges}, in:
  P.~Fraigniaud, Y.~Uno (Eds.), 11th International Conference on Fun with
  Algorithms, {FUN} 2022, May 30 to June 3, 2022, Island of Favignana, Sicily,
  Italy, Vol. 226 of LIPIcs, Schloss Dagstuhl - Leibniz-Zentrum f{\"{u}}r
  Informatik, 2022, pp. 22:1--22:28.
\newblock \href {https://doi.org/10.4230/LIPIcs.FUN.2022.22}
  {\path{doi:10.4230/LIPIcs.FUN.2022.22}}.
\newline\urlprefix\url{https://doi.org/10.4230/LIPIcs.FUN.2022.22}

\bibitem{OUP/oxley11}
J.~Oxley,
  \href{https://doi.org/10.1093/acprof:oso/9780198566946.001.0001}{Matroid
  theory}, 2nd Edition, Vol.~21 of Oxford Graduate Texts in Mathematics, Oxford
  University Press, Oxford, 2011.
\newblock \href {https://doi.org/10.1093/acprof:oso/9780198566946.001.0001}
  {\path{doi:10.1093/acprof:oso/9780198566946.001.0001}}.
\newline\urlprefix\url{https://doi.org/10.1093/acprof:oso/9780198566946.001.0001}

\bibitem{welsh1976matroid}
D.~J.~A. Welsh, Matroid theory, L. M. S. Monographs, No. 8, Academic Press
  [Harcourt Brace Jovanovich, Publishers], London-New York, 1976.

\end{thebibliography}
\end{document}